\documentclass[aps,prl,twocolumn,letterpaper,longbibliography]{revtex4-1}
\usepackage{amssymb}
\usepackage{amsmath}
\usepackage{graphicx}
\usepackage{url}
\usepackage{hyperref}
\newcommand\ket[1]{\ensuremath{|#1\rangle}}

\begin{document}
\title{Creation of a Bose-condensed gas of rubidium 87 by laser cooling}
\author{Jiazhong Hu}\thanks{These two authors contributed equally.}
\author{Alban Urvoy}\thanks{These two authors contributed equally.}
\author{Zachary Vendeiro}
\author{Valentin Cr\'epel}
\author{Wenlan Chen}
\author{Vladan Vuleti\'c}
\affiliation{Department of Physics and Research Laboratory of Electronics, Massachusetts Institute of Technology,
Cambridge, Massachusetts 02139, USA}
\begin{abstract}
We demonstrate direct laser cooling of a gas of rubidium 87 atoms to quantum degeneracy. 
The method does not involve evaporative cooling, is fast, and induces little atom loss. The atoms are trapped in a two-dimensional optical lattice that enables cycles of cloud compression to increase the density, followed by degenerate Raman sideband cooling to decrease the temperature. Light-induced loss at high atomic density is substantially reduced by using far red detuned optical pumping light. Starting with 2000 atoms, we prepare 1400 atoms in 300 ms at quantum degeneracy, as confirmed by the appearance of a bimodal velocity distribution as the system crosses over from a classical gas to a Bose-condensed, interacting one-dimensional gas with a macroscopic population of the quantum ground state. The method should be broadly applicable to many bosonic and fermionic species, and to systems where evaporative cooling is not possible.
\end{abstract}
\maketitle

The ability to prepare quantum degenerate Bose \cite{Anderson198,PhysRevLett.75.1687,PhysRevLett.75.3969} and Fermi \cite{DeMarco1703} gases has opened up a multitude of research areas, including quantum simulation of complex Hamiltonians \cite{quantumsimulation} and quantum phase transitions \cite{PhysRevLett.86.4443,RevModPhys.69.315}. Quantum degenerate gases are prepared in two steps: fast laser cooling until a certain density and temperature limit is reached, followed by slower evaporative cooling to Bose-Einstein condensation (BEC) or below the Fermi temperature. Compared to laser cooling, evaporative cooling \cite{KETTERLE1996181} is slower, requires favorable atomic collision properties (a large ratio of elastic to inelastic collisions), and only a small fraction of the original ensemble is left at the end of the process.
The one exception to this scheme is strontium \cite{PhysRevLett.110.263003}, which features a very narrow optical transition. The narrow line enables the laser cooling of a thermal cloud in a large optical trap, while a small fraction of the ensemble undergoes BEC in a tighter, collisionally coupled trap.

Previous attempts at laser cooling all other species, and especially the alkali workhorse atoms, directly to quantum degeneracy have been limited by light-induced collisions \cite{Walker1994,PhysRevLett.77.1416} that result in substantial heating and loss at the atomic densities required for condensation. With respect to reaching quantum degeneracy, the various laser cooling techniques can be characterized in terms of the attainable phase space density $\mathcal{D}$, which is the peak occupation per quantum state for a thermal cloud. Standard polarization gradient cooling \cite{Dalibard:89} reaches values of $\mathcal{D} \sim 10^{-6}$. A significant improvement of several orders of magnitude is offered by Raman sideband cooling (RSC) \cite{PhysRevLett.80.4149,PhysRevLett.81.5768,PhysRevLett.84.439}, where by keeping the atoms isolated from each other in a three-dimensional (3D) optical lattice, $\mathcal{D} \sim 10^{-2}$ has been reached \cite{PhysRevLett.84.439}. In chromium, a demagnetization cooling technique has also reached $\mathcal D\sim 10^{-2}$ \cite{Ruhrig:15}. Weiss and co-workers pioneered a release-and-retrap compression approach to increase the occupation in a 3D optical lattice \cite{PhysRevLett.82.2262}, and, in combination with RSC, attained a record $\mathcal{D} \sim 0.03$ \cite{PhysRevLett.85.724}, limited by light-induced inelastic collisions in doubly occupied lattice sites. 
 
\begin{figure*}[tp]
\begin{center}
\includegraphics[width=5.3in]{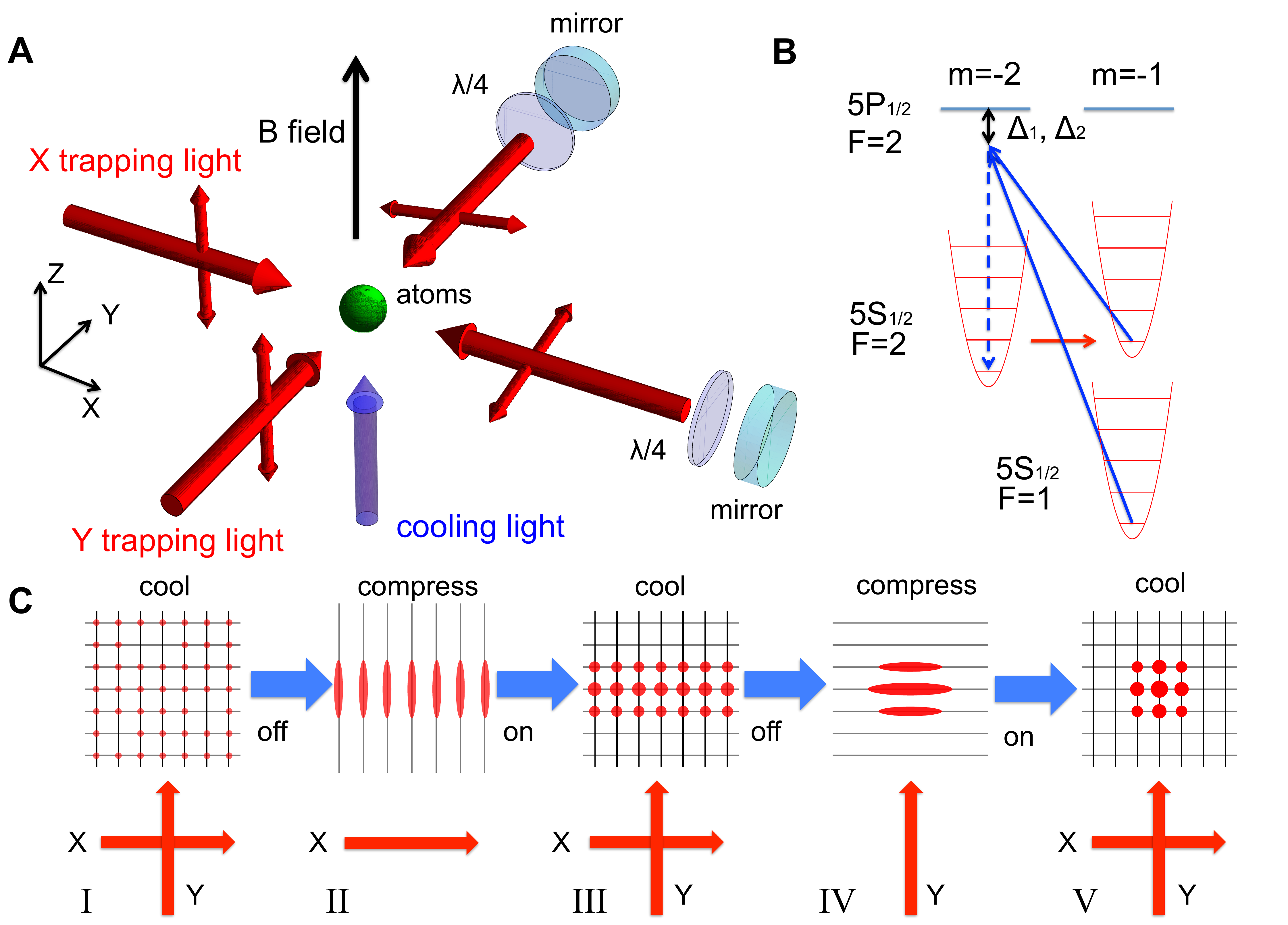}
\end{center}
\caption{Experimental scheme and procedure. \textbf{A} $^{87}$Rb atoms are trapped in a 2D lattice formed by two orthogonal retroreflected trapping beams at 1064~nm. The cooling light at 795-nm propagates along the magnetic field ($z$), and is $\sigma^-$-polarized. \textbf{B} Simplified atomic level structure for dRSC. The Zeeman splitting between two magnetic sublevels is matched to the vibrational splitting in the tightly confined direction. \textbf{C} Release-and-retrap compression sequence used to increase the atomic density. Starting from a sparsely filled 2D lattice, we perform dRSC (I) and then switch off the Y trapping beam to compress the atoms along y in the X trapping beam (II). After a short thermalization time we switch back to the 2D lattice with an increased occupation number per trap (III). The procedure is repeated for the X beam to compress the atoms into a small number of tubes (IV). A final dRSC in this system (V) then yields a condensate.}
\label{fig1}
\end{figure*}

In this Report we show that it is possible to reduce light-induced atom loss and create a Bose condensate by laser cooling $^{87}$Rb, without any evaporation. This is achieved by using far-detuned optical pumping light in degenerate Raman sideband cooling (dRSC) \cite{PhysRevLett.81.5768}. Inelastic atom-atom collisions are further reduced in a 2D lattice geometry with near one-dimensional confinement. We use release-and-retrap compression \cite{PhysRevLett.82.2262,PhysRevLett.85.724} to strongly increase the atomic density after each optical cooling cycle. Starting with 2000 atoms in the central trapping region, we reach quantum degeneracy in 300 ms 
with 1400 atoms. 
In the future, higher trap power should yield significantly larger ensembles, as it is now the limiting factor for the atom number. 

The centerpiece of our setup is a square 2D optical lattice using two retroreflected beams arranged orthogonally to one another, each with a power of 1.1~W. All beams are focussed to an $e^{-2}$ intensity waist of 18~$\mu$m at the position of the atoms. The incoming beams are vertically polarized, while the polarizations of the reflected beams are rotated by $\theta=80^{\circ}$. This induces a polarization gradient in the lattice that provides the required Raman coupling for dRSC \cite{PhysRevLett.81.5768}. For each 1D optical lattice, the trap depth is $U/h=13$~MHz, the axial (tight) vibrational frequency is $\omega_{xy}/(2\pi)=180$~kHz, and the radial vibrational frequency is $\omega_{r\mathrm{2D}}/(2\pi)=4.5$~kHz. 
In the 2D lattice with the two beams combined, this yields $\omega_{z}=\sqrt{2}\omega_{r\mathrm{2D}}=2\pi\times 6.3$~kHz along the weakly confined direction.
A magnetic field $B=0.23$~G, applied along the vertical propagation direction of the optical pumping beam, is set to match the Zeeman splitting between the magnetic sublevels $\ket{F=2,m=-2}$ and $\ket{2,-1}$ to $\hbar \omega_{xy}$.

A cooling cycle consists of a Raman transition $\ket{2,-2}\rightarrow\ket{2,-1}$ induced by the trapping light, that removes one vibrational quantum in the tightly confined direction (Fig.~\ref{fig1}\textbf{B}), followed by optical pumping back to $\ket{2,-2}$. This reduces an atom's motional energy by $\sim \hbar\omega_{xy}$ per optical pumping cycle. The very far detuned trap light that drives the Raman transition (wavelength $\lambda_t=1064$~nm) does not produce any appreciable atom loss, but the optical pumping can induce inelastic binary collisions as an atom pair is excited to a molecular potential that accelerates the atoms before they decay back to the ground state \cite{PhysRevLett.77.1416}. To suppress this process \cite{PhysRevLett.83.943}, we use a $\sigma^-$-polarized optical pumping beam tuned below the $D_1$ line to a local loss minimum in between photoassociation resonances \cite{PhysRevLett.77.1416,Ruhrig:15}. Since an excited atom can also decay to $F=1$, we use bichromatic light with detunings $\Delta_2/(2\pi)=-630$~MHz and $\Delta_1/(2\pi)=-660$~MHz relative to the $\ket{5S_{1/2},F=2}\rightarrow\ket{5P_{1/2},F^{\prime}=2}$ and $\ket{5S_{1/2},F=1}\rightarrow\ket{5P_{1/2},F^{\prime}=2}$ transitions, respectively. This modified optical pumping configuration represents the most significant improvement over earlier schemes \cite{PhysRevLett.85.724,PhysRevLett.84.439,PhysRevLett.81.5768,PhysRevLett.80.4149}, and we find that light-induced inelastic collisions are suppressed by at least an order of magnitude.

\begin{figure}[ht!]
\begin{center}
\includegraphics[width=4.2in]{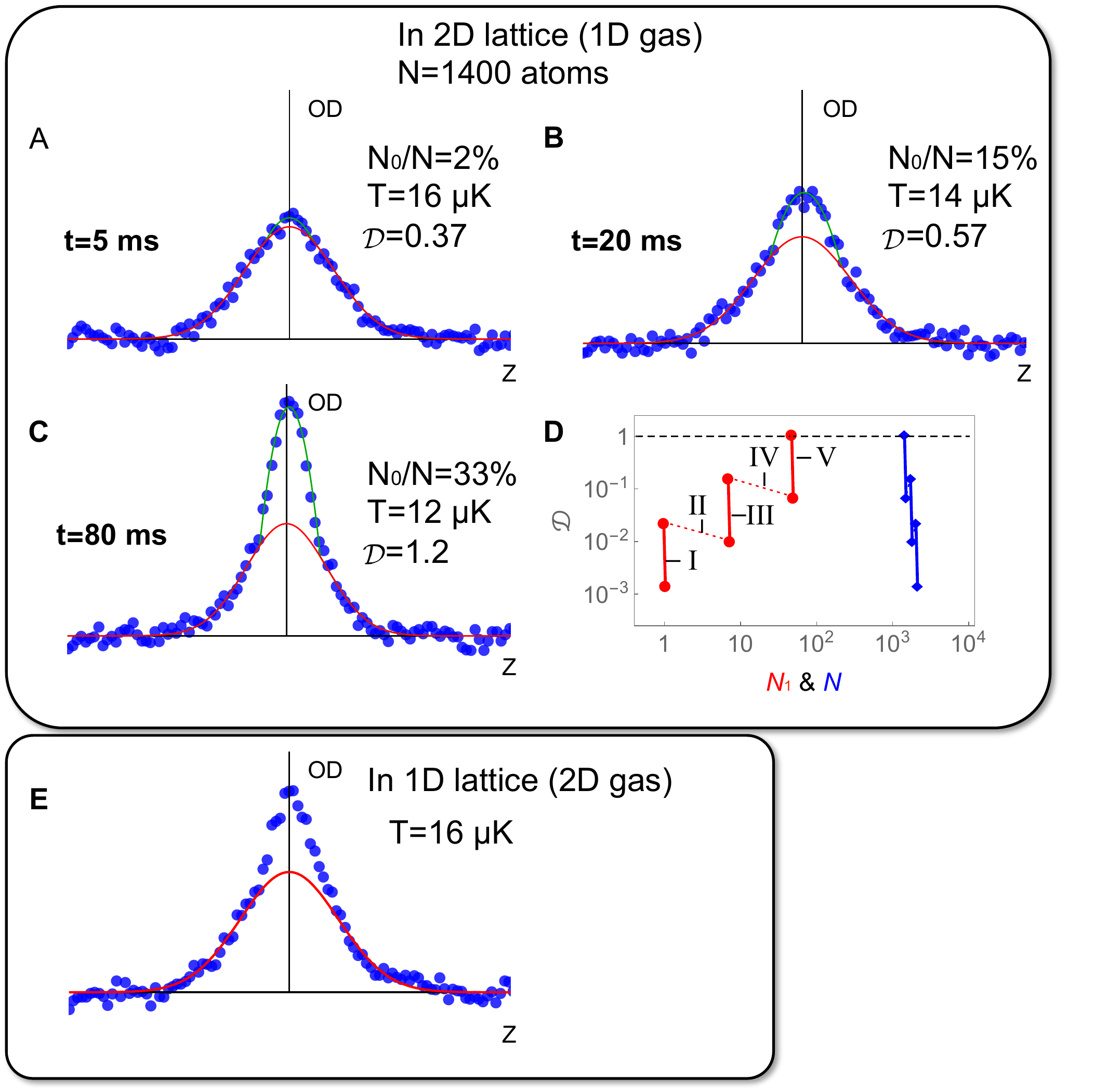}
\end{center}
\caption{\textbf{A} to \textbf{C}: Bimodal velocity distribution emerging during the final laser cooling stage indicating macroscopic population of the ground state. Observed optical depth (OD) along $z$ after a ballistic expansion of 1.3~ms for cooling times of 5~ms (\textbf{A}), 20~ms (\textbf{B}) and 80~ms (\textbf{C}), averaged over 200 repetitions of the experiment. 
The red lines show Gaussian fits to the wings of the distributions, and green lines are quadratic fits to the remaining distribution in the center. 
Here the intensity of the trapping beams is ramped down in 400~$\mu$s, slowly with regard to the axial trapping frequency but rather quickly with regard to the motion along $z$, in order to reduce the interaction energy. \textbf{D}: Evolution of the total atom number $N$ (blue) or atom number per lattice tube $N_1$ (red), vs the phase space density $\mathcal D$ during the sequence, with the steps labeled I to V as defined in Fig.~\ref{fig1}\textbf{B}. The solid lines represent the dRSC process and the dotted lines represent the spatial compression. Each cooling step enhances $\mathcal D$ by one order of magnitude, then the release-and-retrap compression increases the peak occupation number $N_1$ while slightly decreasing $\mathcal D$.
\textbf{E}: Velocity distribution along $z$ for the same parameters as in \textbf{C}, but observed for a 2D gas after releasing the atoms into the 1D lattice Y.}
\label{fig2}
\end{figure}

\begin{figure}[ht!]
\begin{center}
\includegraphics[width=2.5in]{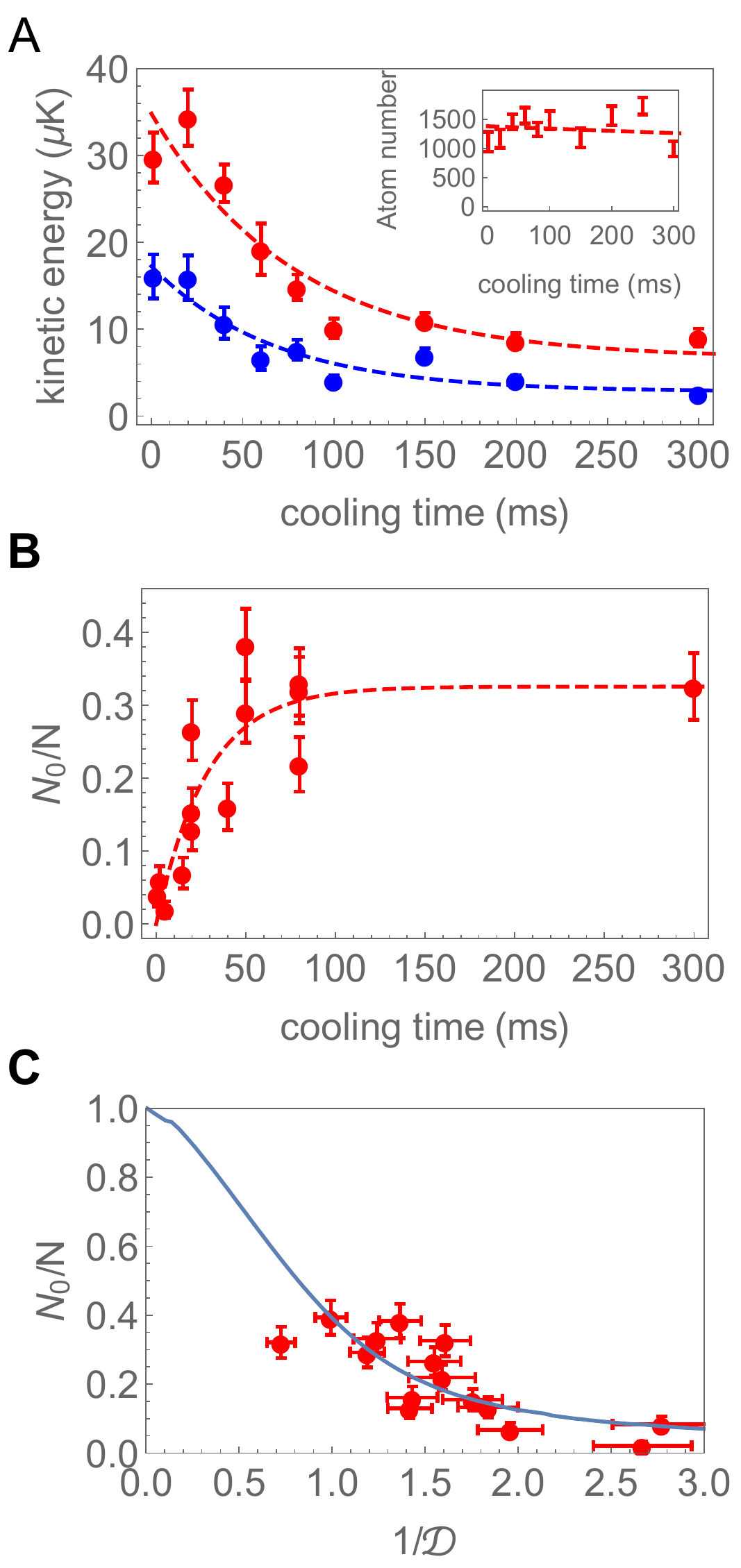}
\end{center}
\caption{\textbf{A} Cooling performance in the final stage. Average kinetic energy in the directions of strong (blue) and weak (red) confinement against cooling time. After 80~ms of cooling, the atoms are in the vibrational ground state of the axial direction. The inset shows that there is almost no atom loss during the final cooling stage. The loss suppression is due to far-red-detuned optical pumping and the 1D geometry. \textbf{B} Condensate fraction vs cooling time. In \textbf{A} and \textbf{B} the dashed lines are exponential fits shown as a guide to the eye. \textbf{C} Condensate fraction vs the inverse peak phase space density $\mathcal{D}^{-1}$. The blue line is the theoretical prediction for an ideal gas of 50 atoms in a 1D trap \cite{PhysRevA.54.656}, shown as a guide to the eye.}
\label{fig3}
\end{figure}

The experimental sequence starts by accumulating cold $^{87}$Rb atoms in a magneto-optical trap, loading them into the 2D lattice using polarization gradient cooling \cite{PhysRevLett.82.2262}, and cooling them for 100~ms with dRSC. This prepares the atoms near the vibrational ground state in the strongly confined $x$ and $y$ directions (the kinetic energy upon instantaneous trap release is $K_{xy}/h=$50~kHz, close to $\frac{1}{4} \omega_{xy}/(2\pi)=45$~kHz), while in the vertical direction ($z$) the atoms are cooled to $T_z\approx 12$~$\mu$K ($K_z/h=120$~kHz) via collisional thermalization between the axial and radial directions of the tubes. At this point there are $N=2000$ atoms in the 2D lattice with a peak occupation of $N_1\approx 1$ atom per tube, corresponding to a peak phase space density $\mathcal{D}=0.02$, and peak density $n_0=2.2\times 10^{14}$~cm$^{-3}$. In order to further increase the density and $\mathcal{D}$, we apply release-and-retrap compression \cite{PhysRevLett.82.2262} by adiabatically turning off (in 400~$\mu$s) the Y trapping beam, such that the cloud shrinks in the $y$ direction due to the radial confinement of the X beam (Fig.~\ref{fig1}\textbf{C}). After thermalization for 10~ms, the spatial extent of the cloud can be estimated by $z=\sqrt{k_B T_z/m}/\omega_{r\mathrm{2D}} = 1.1\,\mu$m, where $T_z=10$ $\mu$K is the measured radial temperature, $m$ is the atomic mass of $^{87}$Rb, and $k_B$ is Boltzmann's constant. The lattice beam is then turned back on in 1~ms. This loads the compressed ensemble back into a 2D lattice, resulting in a higher temperature ($T\sim 50$ $\mu$K), and we apply again dRSC for 100~ms. This yields again $K_{xy}\approx \hbar\omega_{xy}/4$, $T_z=$12~$\mu$K, but at a peak occupation number of $N_1=6.9$ atoms per tube for a total atom number $N=1700$. We repeat this procedure for the X lattice beam and end up with $N=1400$ and $N_1=47$ at a peak density $n_0=1\times 10^{16}$ cm$^{-3}$. At this point the trapped atom ensemble is below our optical resolution limit of 8~$\mu$m, and $N_1$ is estimated from the measured temperatures in the corresponding 1D lattices, and the separately measured trap vibration frequencies. Fig.~\ref{fig2}\textbf{D} shows the evolution of $N$, $N_1$ and $\mathcal D$ during the sequence that brings the system close to $\mathcal D=1$. We would like to emphasize that evaporation is not occurring at any point, since temperature reduction is only observed when the cooling light is on, and $k_BT\le 0.1 U$ at all times. 

When we then apply the final dRSC stage for up to 100~ms, we observe the gradual appearance of a characteristic signature of condensate formation, a bimodal velocity distribution along the $z$ direction that becomes more pronounced with longer dRSC time $t$ (Fig.~\ref{fig2}). If we attempt to fit the observed distribution for $t=80$~ms with a single Gaussian curve, the reduced $\chi^2$ is $\chi^2=137$ (Fig.~\ref{fig2}\textbf{C}), as opposed to $\chi^2=0.91$ for a bimodal distribution with a parabolic density component in the center. (At $t=5$~ms, the corresponding values are $\chi^2=0.89$ and $\chi^2=0.75$, respectively.) The bimodal distribution persists if we adiabatically turn off the X trap after cooling for $t=80$~ms and observe the 2D gas in a 1D lattice (Fig.~\ref{fig2}\textbf{E}).

In Fig.~\ref{fig3}, we show the evolution of the kinetic energy, the condensate fraction, and the atom number as a function of cooling time for the final cooling stage. Along the tightly confined direction, we reach $K_{xy}/h=50$~kHz $=1.1\times\frac{1}{4}\omega_{xy}/(2\pi)$, indicating cooling to the vibrational ground state. Along the vertical, weakly confined direction, we reach an average kinetic energy $K_z/h= 120$~kHz. Additionally, we observe only very limited atom loss ($<5\%$ at a peak density of $n_0=1\times10^{16}$ cm$^{-3}$, inset to Fig.~\ref{fig3}\textbf{A}), confirming that light-induced losses are strongly suppressed. (For the initial two cooling stages, we observe a similar temperature evolution at a slightly smaller loss.) Fig.~\ref{fig3}\textbf{C} also shows the condensate fraction $N_0/N$, defined as the fractional area under the narrower peak in the bimodal distribution, vs the calculated inverse phase space density $1/\mathcal D$ (see supplementary materials). Within an estimated systematic error of a factor of 2 for $\mathcal D$, the onset of the bimodal distribution is observed near $\mathcal D=0.7$. Note that in 1D systems, only a smooth crossover to a quantum degenerate gas occurs \cite{PhysRevLett.85.3745,PhysRevA.75.031606}, which is in agreement with our observation for the condensate fraction $N_0/N$. 
For our parameters the system is at the same time in the crossover region between a weakly interacting 1D gas \cite{PhysRevLett.105.265302} and the strongly interacting Tonks-Girardeau gas \cite{Kinoshita2004,Paredes2004} (the calculated dimensionless interaction parameter is $\gamma\approx 2.7$ at the peak local density), as well as in the crossover region between a 1D Bose gas and a 3D finite-size condensate \cite{PhysRevLett.87.130402} (see supplementary materials). While the exact character of the condensate is therefore ambiguous, the velocity distribution (Fig.~\ref{fig2}) clearly reveals a macroscopic population of the ground state at quantum degeneracy.

Since the atomic cloud is below our optical resolution, the atomic density cannot be directly determined through optical imaging. However, an independent verification is possible by measuring 3-body loss, where the loss coefficient ($K=2.2\times 10^{-29}$~cm$^6$s$^{-1}$ for a 3D thermal gas \cite{threebody}) has been previously determined. It is known that 3-body loss is strongly suppressed in a 1D gas for $\gamma\gtrsim 1$ \cite{PhysRevLett.92.190401,PhysRevLett.107.230404}, and indeed we do not detect any loss in the 1D tubes. Instead, we measure the 3-body recombination in the 1D lattice, i.e. for a 2D gas, where we observe a three-body recombination timescale of 300~ms, from which we determine a peak density of $5.3\times 10^{14}$~cm$^{-3}$ in the 2D gas, corresponding to a peak atom number of $N_1\ge 45$ atoms per tube for the 1D ensembles. This is in agreement with the previous estimation, and yields a phase space density $\mathcal D=N_1(\hbar \omega_{z}/k_B T)(\hbar\omega_{xy}/4 K_{xy})=1.1$ at the onset of quantum degeneracy.

In conclusion, we have directly laser cooled a gas of alkali atoms to quantum degeneracy, which had remained an elusive goal since the early quest for BEC. We expect that the atom number can be substantially increased in the future using higher trap power, and that the method can be applied to various bosonic as well as fermionic atomic species, potentially even under conditions where evaporative cooling is impossible. The far-detuned optical pumping light may also enable atom-number resolving measurements in quantum gas microscopes \cite{microscope1,microscope2}. Finally, the fast preparation may pave the way for further studies of the nature and coherence properties of the condensate at the boundary between 3D and 1D, and into the strongly correlated Tonks gas regime \cite{Kinoshita2004,Paredes2004}.

This work was supported by the NSF, NSF CUA, NASA and MURI grants through AFOSR and ARO. The authors gratefully acknowledge stimulating discussions with Cheng Chin, Wolfgang Ketterle, Robert McConnell and Martin Zwierlein.
\bibliographystyle{apsrev4-1}
\bibliography{BEC}

\clearpage
\widetext
\begin{center}
\textbf{\large Supplementary materials: Raman sideband cooling of rubidium 87 to Bose-Einstein condensation}
\end{center}

\setcounter{equation}{0}
\setcounter{figure}{0}
\setcounter{table}{0}
\setcounter{page}{1}
\makeatletter
\renewcommand{\theequation}{S\arabic{equation}}
\renewcommand{\thefigure}{S\arabic{figure}}
\renewcommand{\bibnumfmt}[1]{[S#1]}
\renewcommand{\citenumfont}[1]{#1}

\subsection{Experimental details}
The two trapping beams differ in frequency by 160~MHz to avoid interference effects and their powers can be independently controlled by separate acousto-optical modulators. The intensity of the optical pumping beam at 795~nm is set to an off-resonant scattering rate $\Gamma_s\sim 2\times10^3$ s$^{-1}$, and the power ratio between the two frequency component is set to have a 3 times stronger scattering rate on the $\ket{5S_{1/2},F=1}\rightarrow\ket{5P_{1/2},F=2}$ transition. This prevents the atoms that have decayed to the $\ket{F=1}$ state from undergoing heating Raman transitions. 
The atoms are imaged after a time-of-flight of typically 1.3~ms via absorption imaging on the cycling transition of the D2-line, in the plane defined by the lattice beams, and at a 20$^\circ$ angle relative to the X lattice beams (see Fig.~\ref{fig1} of the main text). 
We summarize all relevant experimental parameters in Table~\ref{tab6.1}.

\begin{table}[b]
	\begin{center}
	\caption{Experimental parameters.}
	\begin{tabular}{|c|r|}
	\hline
	trap wavelength $\lambda$			&1064 nm	\\ \hline
	power of each trapping beam			&1.1 W                              \\ \hline
	waist								&18 $\mu$m \\ \hline
	$\omega_{xy}/(2\pi)$				&180 kHz \\ \hline
	$\omega_{r2\textrm{D}}/(2\pi)$		&4.5 kHz \\ \hline
	$\omega_z/(2\pi)$					&6.3 kHz \\ \hline
	trap depth in the 1D lattice $U$/h	&13 MHz \\ \hline
	magnetic field $B$					&0.23 G \\ \hline
	$\Delta_1/(2\pi)$					&$-660$ MHz \\ \hline
	$\Delta_2/(2\pi)$					&$-630$ MHz \\ \hline
	\end{tabular}
	\label{tab6.1}
	\end{center}
\end{table}

\subsection{Estimation of the phase space density $\mathcal D$}
We measure the kinetic energies $K_{xy}$ and $K_{z}$, and compare them to the trapping vibrational frequencies $\omega_{xy}$ and $\omega_z$. Hence, we estimate the relative ground state occupation along each direction when the chemical potential is zero. Assuming $T_\beta$ ($\beta=x$, $y$ or $z$) is the temperature along direction $\beta$ with the vibrational frequency $\omega_\beta$, the kinetic energy $K_\beta$ is related to $T_\beta$ by
\begin{equation}
K_\beta={1\over 4}\hbar\omega+{1\over 2}\hbar\omega{1\over e^{\hbar\omega_\beta\over k_B T_\beta}-1}.
\end{equation} 
Then, we know the relative ground state occupation is
\begin{equation}
p_{0,\beta}=1-e^{-{\hbar\omega_\beta\over k_B T\beta}}={2\over {4K_\beta\over \hbar\omega_\beta}+1}.
\end{equation}
The occupation of the 3D ground state is
\begin{equation}
P_0=p_{0,x}p_{0,y}p_{0,z}={2\over{4K_z\over\hbar \omega_z}+1 }\left({2\over{4K_{xy}\over \hbar\omega_{xy}}+1 }\right)^2.
\end{equation}
Thus the phase space density $\mathcal D$ is calculated as
\begin{equation}
\mathcal{D}=N_1 P_0=N_1 {2\over{4K_z\over\hbar\omega_z}+1 }\left({2\over{4K_{xy}\over\hbar\omega_{xy}}+1 }\right)^2\approx N{\hbar \omega_z\over k_B T_z}\left({2\over{4K_{xy}\over\hbar\omega_{xy}}+1 }\right)^2,
\end{equation}
here $T_z=2 K_z/k_B \gg \hbar\omega_z/k_B$ is the measured temperature along $z$. $K_x=K_y=K_{xy}$ is the measured kinetic energy along $x$ or $y$, and $N_1$ is the peak atom number per lattice tube.

\subsection{Three-body recombination measurement}
\begin{figure}[htbp]
\begin{center}
\includegraphics[width=4.5in]{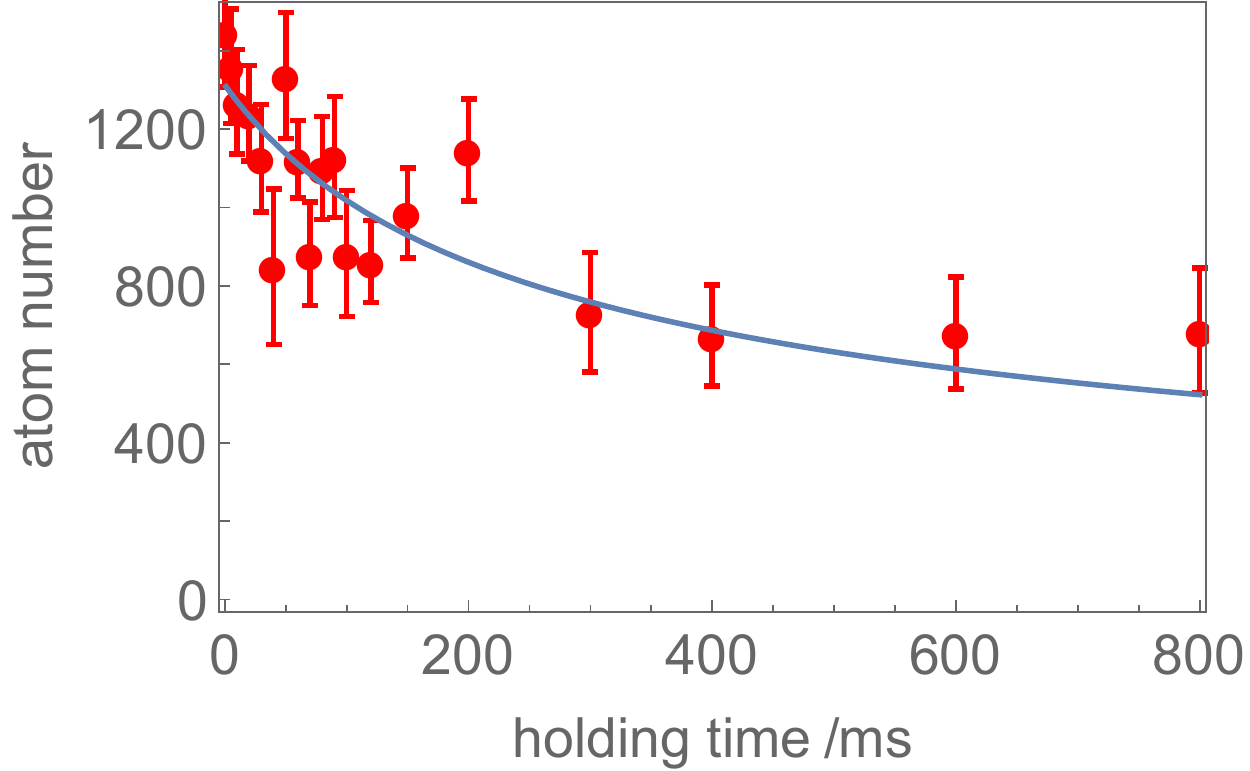}
\end{center}
\caption{Atom number in the $\ket{F=2}$ state as a function of the holding time in the Y lattice. The blue solid line is a fit to the analytic solution of the three-body loss decay, with an initial loss timescale of $\tau=300$~ms.}
\label{sfig1}
\end{figure}

It was previously reported that three-body losses are strongly reduced in a one-dimensional cloud \cite{PhysRevLett.92.190401,PhysRevLett.107.230404}. Indeed we are not able to measure any significant three-body loss in the tubes. Therefore we measured the three-body recombination rate in a two-dimensional geometry by transferring the atoms to the Y optical lattice (i.e. by merging the tubes in the x direction) after the final cooling for 10 ms, and measuring the atom number as a function of the holding time (Fig.~\ref{sfig1}).

By fitting the theoretical model for the three-body loss, we obtain an initial loss timescale of $\tau=300$~ms. Assuming a thermal velocity distribution for the atoms and averaging over the Gaussian density profile in each trap and over the different lattice sites in the 1D lattice, we obtain the following relation between the initial peak density in the 2D gas and $\tau$: 
\begin{equation}
n=\left(\frac{3}{2 \tau K}\right)^{1/2}=5\times 10^{14} \quad \mathrm{cm}^{-3}.
\end{equation}
Here we used the value $K=2.2\times 10^{-29}$~cm$^6$s$^{-1}$ from \cite{threebody} as the three-body loss coefficient for the $\ket{F=2}$ state for a classical (non-condensed) gas. This density is also consistent with the peak atom number per tube derived from the atomic temperature and trap vibrational frequencies.

\subsection{Detuning dependence of the cooling sequence}
\begin{figure}[htbp]
\begin{center}
\includegraphics[width=3in]{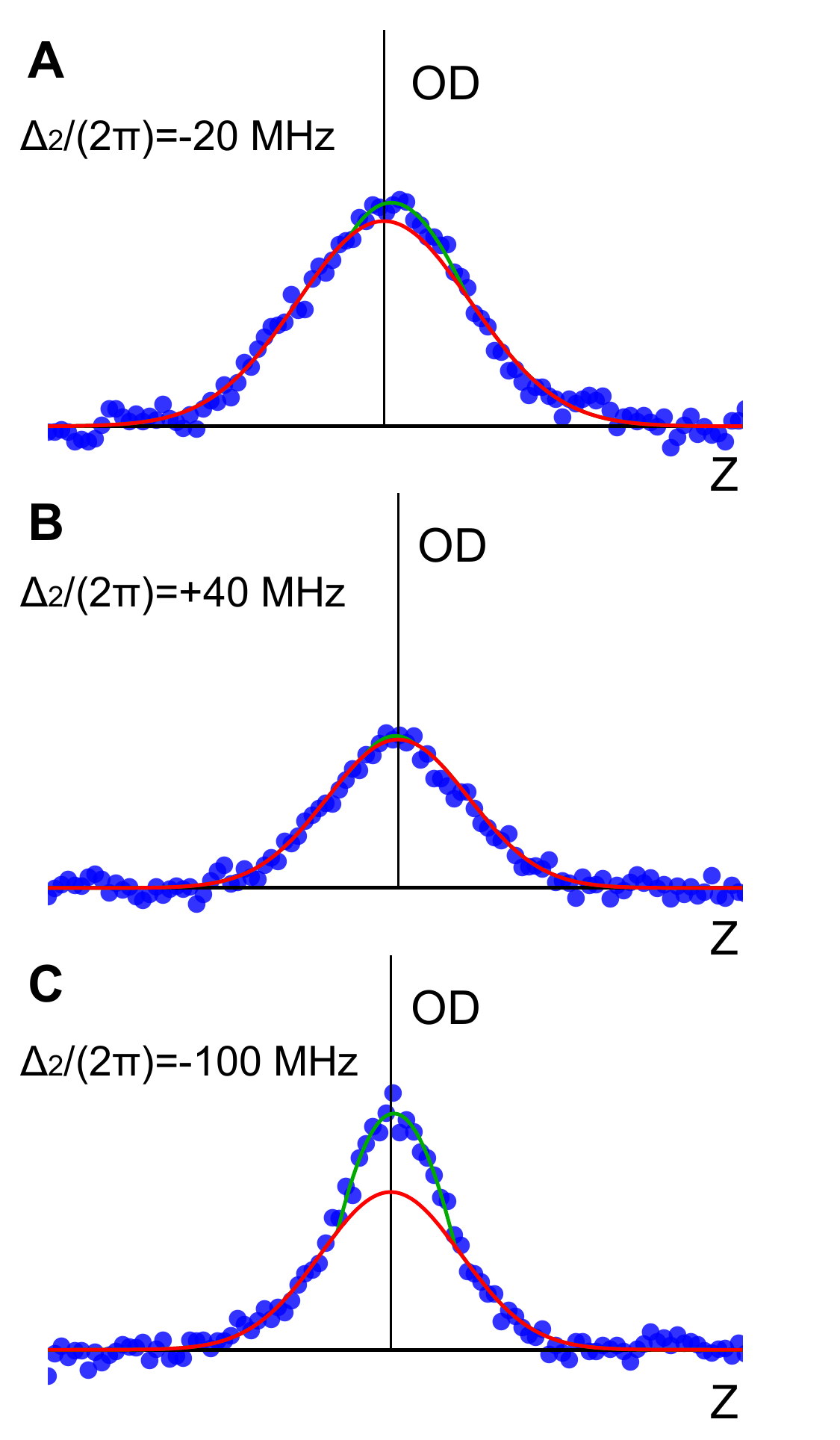}
\end{center}
\caption{Velocity distribution of the atoms along the Z direction after 80~ms of cooling in the final cooling stage, for various detunings of the cooling beam $\Delta_2/(2\pi)=-20$~MHz (\textbf{A}), $\Delta_2/(2\pi)=+40$~MHz (\textbf{B}) and $\Delta_2/(2\pi)=-100$~MHz (\textbf{C}), avaraged over 100 repetitions. The data is fitted with a bimodal distribution in the same way as in Fig.~2 of the main text.}
\label{sfig2}
\end{figure}

We also test a few different detuning for the optical pumping beam of dRSC. 
For each detuning setting, the laser intensity was adjusted to maintain a scattering rate of $\Gamma_s\sim 2\times 10^3$~s$^{-1}$. 
When using exactly the same cooling sequence as described in the main text, we succeed to produce a condensate also at a the smaller detuning $\Delta_2/(2\pi)=-100$~MHz (Fig.~\ref{sfig2}\textbf{C}) but not at $\Delta_2/(2\pi)=-20$~MHz or $+40$~MHz (Fig.~\ref{sfig2}\textbf{A}-\textbf{B}). 

\subsection{BEC regime}
For our parameters, at the critical temperature for quantum degeneracy of $k_B T\approx \hbar \omega_{xy}$, the system is at the boundary between a 3D gas and a 1D gas \cite{PhysRevLett.87.130402}. Furthermore, the dimensionless interaction parameter $\gamma=m g_1/\hbar^2 n_1$ \cite{PhysRevLett.85.3745}, where $n_1$ is the 1D density, and $g_1\sim 2\hbar \omega_{xy} a$ is the interaction strength for the 3D scattering length $a$, for our system is $\gamma\approx 2.7$ at the peak 1D density. This means that the system is also at the boundary between a weakly interacting Thomas-Fermi gas ($\gamma\ll 1$, for high linear density $n_1$) and a strongly correlated Tonks gas ($\gamma\gg 1$, for low $n_1$) \cite{PhysRevLett.85.3745,Kinoshita2004,Paredes2004}. In fact, the latter has been measured to exhibit substantially lower collisional loss due to the reduced correlation function \cite{PhysRevLett.92.190401,PhysRevLett.107.230404}. This effect may also help further reduce the light-induced loss during dRSC in the near-1D geometry.

\end{document}